\documentclass{cpbtex}
\usepackage{times}
\usepackage{CJK}
\usepackage{titlesec}
\usepackage{amsthm,amssymb}
\usepackage{color}
\usepackage{graphicx}
\usepackage{dcolumn}
\usepackage{bm}
\usepackage{amsthm}
\usepackage{ulem}
\usepackage{titlesec}
\usepackage{graphicx}
\usepackage{colortbl}
\usepackage{listings}
\usepackage{makecell}
\usepackage{indentfirst}
\usepackage{fancyhdr}
\usepackage{setspace} 
\usepackage{bm}

\usepackage{array}
\usepackage{multirow}
\usepackage{tabularx}
\usepackage{booktabs}

\newcolumntype{Y}{>{\centering\arraybackslash}X}

\usepackage{color}
\usepackage{xfrac}
\usepackage{graphicx}
\usepackage{dcolumn}
\usepackage{bm}
\usepackage{amsthm}
\usepackage{cases}
\usepackage{txfonts}

\newcommand{\bx}{\mathbf{x}}
\newcommand{\bv}{\mathbf{v}}
\newcommand{\bE}{\mathbf{E}}
\newcommand{\bP}{\mathbf{P}}
\newcommand{\bp}{\mathbf{p}}
\newcommand{\rp}{\mathrm{p}}
\newcommand{\rb}{\mathrm{b}}
\newcommand{\rM}{\mathrm{M}_\mathrm{a}}
\newcommand{\bk}{\mathbf{k}}
\newcommand{\dif}{\mathrm{d}}
\newcommand{\me}{\mathrm{e}}
\newcommand{\mi}{\mathrm{i}}

\newcommand{\tx}{\tilde{x}}
\newcommand{\tv}{\tilde{v}}
\newcommand{\teta}{\tilde{\eta}}
\newcommand{\tvx}{\tilde{v}_x}
\newcommand{\ttt}{\tilde{t}}
\newcommand{\tphi}{\tilde{\phi}}
\newcommand{\bphi}{\bar{\phi}}
\newcommand{\tT}{\tilde{T}}

\newcommand{\tN}{\tilde{N}}
\newcommand{\tn}{\tilde{n}}
\newcommand{\tu}{\tilde{u}}
\newcommand{\ttauu}{\tilde{\tau}}
\newcommand{\hp}{\hat{p}}
\newcommand{\neo}{n_{\me 0}}
\newcommand{\nee}{n_{\me}}
\newcommand{\kB}{k_\mathrm{B}}
\newcommand{\kp}{k_\mathrm{p}}
\newcommand{\Te}{T_\me}
\newcommand{\vp}{v_\rp}
\newcommand{\tlambdae}{\tilde{\lambda}_\me}
\newcommand{\tOmegai}{\tilde{\Omega}_\text{p}}

\newcommand\eq[1]{\begin{align} #1 \end{align}}
\newcommand\Eq[1]{Eq.~\eqref{#1}}

\newcommand{\altfrac}[2]{\ifmmode\def\tmp{$}\else\def\tmp{}\fi\mbox{%
    {\raisebox{.24\ht\strutbox}{\tmp#1\tmp}}%
    \kern-2.2pt\scalebox{1.6}[1.5]{/}\kern-1.8pt%
    {\tmp#2\tmp}%
    }}

\begin{document}

\title{Primordial electric fields before recombination in the early Universe}


\author{Xi-Bin Li \thanks{Corresponding author. E-mail:lxbimnu@mail.bnu.edu.cn}\\
{$^1$College of Physics and Electronic Information, Inner Mongolia Normal University, }\\{81 Zhaowuda Road, Hohhot, 010022, Inner Mongolia, China}\\
{$^2$Inner Mongolia Key Laboratory for Physics and Chemistry of Functional Materials, }\\{Inner Mongolia Normal University,
        81 Zhaowuda Road, Hohhot, 010022, Inner Mongolia, China}\\
{$^3$School of Physics and Technology, Wuhan University, 430072, Wuhan, China}}

\maketitle


\begin{abstract}
This work is a supplement on the previous research about primordial electromagnetic fields.
In this work, three important problems are discussed: the evolution of primordial electric fields, the electric and particle densities' solitons in plasma before recombination and
their influences on the power spectra of cosmic microwave background. Detailed computations show that the primordial electric fields dissipate by Landau damping effect
on both large scale and small scale and there is no impact on the spectrum. While, before recombination, there exist solitary waves stably propagating in plasma whose speed is
significantly slower than that of baryonic acoustic oscillations, working only at extremely small scale. On the other hand, the amplitude of solitons is
so weak that only a significantly small contribution on the phase of baryon acoustic oscillations, so there merely exist the messages about such electric solitary waves on the spectrum.
In a word, as relevant monographs on cosmology, neglecting the electromagnetic fields (electric fields at least) is a reasonable treatment on the calculations of cosmic microwave background.
However, the protonic density fluctuations show a form of KdV equation while its propagation as a stable solitary wave, leading a probability to the origin of fluctuation
promoting the generation and evolution of galaxies.
\end{abstract}



\section{Introduction} \label{introduction}

Primordial electromagnetic fields are often considered as the essential components in early Universe \cite{PhysRevD.52.6694,RevModPhys.74.775}. They relate to the
electromagnetic dynamical process in the primordial plasma, cosmological perturbation dynamics and galaxies' generation and evolution \cite{RevModPhys.74.775,Subramanian_2016,GRASSO2001163}.
Precious studies focus mainly on the primordial magnetic problems. The origin of the magnetic fields have been widely studied and may generate
in different ways, like inflation \cite{PhysRevD.83.023514,PhysRevD.62.103512,PhysRevD.55.7398}, quark confinement \cite{PhysRevD.55.4582,PhysRevLett.85.5268}
and neutrino decoupling \cite{PhysRevD.46.5346,PhysRevD.57.7139,PhysRevD.81.043517}. Such a field plays an important role in the electrodynamic properties \cite{PhysRevD.52.6694},
baryon acoustic oscillations \cite{ADAMS1996253,Cai_2010} and tensor perturbations \cite{PhysRevD.69.063006,PhysRevD.83.123003,10.1111/j.1365-2966.2009.14727.x}.
It has attracted a lot of interests on the study of the tensor perturbed mode
induced by the cosmic magnetic fields, since it they give rise to a anisotropy signal on B-mode polarizations of cosmic microwave background (CMB)
\cite{cmb1,cmb2,cmb3,cmb4,cmb5,cmb6,cmb7}.
However, the average amplitude of magnetic field should be small enough to be consistent with the observations on spatial large-scale isotropy.
This is the reason why linear nonequilibrium or instability method is applied to study the perturbation dynamics
\cite{Subramanian_2016,PhysRevD.100.023524,Hutschenreuter_2018,Chowdhury_2019,per1,per2,per3}.

But unfortunately, the importance of the study on primordial cosmological electric fields has not been realized till now. It may be partly because the previous work
did not show any amazing results about electric fields or because the attractive from electric effects in primordial plasma are not so strong compared with it from magnetic effects.
But it needs to point out, in plasmonics, the electric properties have an equally important position as magnetic ones \cite{pecseli2020waves,doi:10.1142/11168}.

Vlasov equation, as the mathematical tool on the plasma statistical properties in electromagnetic fields, is deeply studies during the last century and widely
used in the nuclear industry and other fields \cite{PhysRevD.58.125015,sedlacek_1971,clark2005nuclear,doi:10.1063/1.1706871,Vlasov1,Vlasov2}.
Relevant theoretical results, like Landau damping effect, acoustic waves and electric solitary wave, have been
solidly tested by modern experiments. Therefore it's reasonable to deduce that the corresponding phenomena should also occur in cosmic primordial plasma.
Now the extension on relevant results to the early Universe is attempted
and several questions not discussed earlier are going to propose in this paper. The first question is the generation of magnetic field
must be accompanied with the generation of electric field. So it's necessary to analyse the evolution of electric fields and discuss how to propagate in the plasma.
The second question is that whether there is a special way to persist the nonlinear perturbation after recombination, especially the protonic density fluctuation,
since perturbations continue to grow up in the absence of optical pressure during and after recombination. The last question is whether primordial cosmological electric fields influence
on the CMB power spectra and how significantly they are.

In this paper, I mainly concentrate on answering these three problems. The method, of course, is the Vlasov equation, which is a Boltzmann equation describing a charged particle's state in
phase space moving in the electromagnetic field. The optimal scenario to obtain the power spectra of CMB is the Boltzmann equation, as illustrated
in series of monographs \cite{2003moco.book.....D,giovannini2008primer,durrer2021cosmic}.
So a potentially feasible method to study the effects on perturbed dynamics, evolutions for large-scale structure and CMB from the electromagnetic fields
is the Vlasov equation. As the critical part of this work, the first step is obtaining the Vlasov equation in perturbed FRW metric which could be accomplished by applying
the geodesic equation. Based on this equation, the problems introduced previous could solved via the existing mature solutions. Relevant conclusions are quite brief but persuasive.
The initial electric fields dissipate out via Landau damping, intrinsic electric solitary waves propagate with a speed something like average thermal velocity but not exactly and
there is almost no influences on CMB power spectra except at extremely small scale.

The following contents of this paper is organized as follows. In Sec.~\ref{Vlasov}, the Vlasov equation in perturbed FRW Universe is obtained via the geodesic equation, which is the
key step of this work. In next section, it's proofed the initial electric fields generated before recombination are gong to dissipate, during which it is assumed the electrons state under a thermal equilibrium
and obey the Maxwell distribution. In Sec.~\ref{soliton}, the existence of electric solitary waves in primordial plasma is proofed via a series of nonlinear equations,
which could retain after recombination to help the generations and evolutions of the galaxies and galaxy cluster. In Sec.~\ref{ASS}, the relation
between the acoustic oscillations and cosmic electric fields is analytically computed, and the influences on CMB are also discussed.
In the final section, brief conclusions and further research programmes are given.

\section{\label{Vlasov}Vlasov Equation in Perturbed FRW Universe}                                       

The scaler perturbation in perturbed FRW Universe is described by the metric
\eq{
    g_{00}(\bx,t) &= -1-2\Psi(\bx,t), \nonumber \\
    g_{0i}(\bx,t) &= 0,  \nonumber \\
    g_{ij}(\bx,t) &= a^2\delta_{ij}[1+2\Phi(\bx,t)], \label{metric}
}
where $a$ is scale factor dependent only on cosmic time and $\Psi$, $\Phi$ are two variables describing the scaler perturbation under conformal gauge.
The distribution function of a particle under equilibrium state depends on position coordinate $\bx$, momentum $\bp$ and time $t$ whose complete differential on $t$ reads
\eq{
    \frac{\dif f}{\dif t} = \frac{\partial f}{\partial t}+\frac{\partial f}{\partial x^i}\frac{\partial x^i}{\partial t}+\frac{\partial f}{\partial p}\frac{\partial p}{\partial t}
        +\frac{\partial f}{\partial \hp^i}\frac{\partial \hp^i}{\partial t}. \label{df}
}
The four-momentum is defined as
\eq{
    P^\mu = m U^\mu \label{4P}
}
which relats to the four-velocity
\eq{
    U^\mu = \frac{\dif x^\mu}{\dif \lambda}, \label{4U}
}
where $m$ is the rest mass of a particle and $\lambda$ is the proper time. The second term on the right-hand in \Eq{df} could be reexpress in terms of four-momentum
by applying the definitions in \Eq{4P}. Therefor,
\eq{
    \frac{\dif x^i}{\dif t}=\frac{\dif x^i}{\dif \lambda}\frac{\dif \lambda}{\dif t}=\frac{P^i}{P^0}. \label{dxidt}
}
The four-momentum of a massive particle is also derived in terms of particle's energy and momentum:
\eq{
    P^\mu = \left[E(1-\Psi),p\hp^i\frac{1-\Phi}{a}\right], \label{4P1}
}
where $p^2=g_{ij}P^i P^j$. Similarly, it's also available to get the complete differential of a massive particle's, like electron, distribution function
\eq{
    \frac{\dif f_\me}{\dif t} = \frac{\partial f_\me}{\partial t}+\frac{\partial f_\me}{\partial x^i}\frac{\partial x^i}{\partial t}+\frac{\partial f_\me}{\partial E}\frac{\partial E}{\partial t}, \label{dfe}
}
where the term disappears in \Eq{df} with the derivative respect to orientation vector of momentum $\hp^i$. The electrodynamics equation in curved space-time is written as
\eq{
    \frac{\mathrm{D}P^\mu}{\mathrm{D}\lambda} = \frac{q}{c}F^{\mu\nu}U_\nu, \label{electrodynamics1}
}
with $F^{\mu\nu}$ denoting the electromagnetic tensor. Setting $\mu=0$, \Eq{electrodynamics1} describing a electron's motion is rewritten in terms of Christoffel symbol as
\eq{
    m\frac{\dif U^0}{\dif \lambda}+m\Gamma^{0}_{\ \alpha\beta}P^\alpha P^\beta = -\me_0E^i U_i. \label{electrodynamics2}
}
Repeating the computations in Ref.~\cite{2003moco.book.....D}, the derivative of energy to cosmic time gives
\eq{
    \frac{\dif E}{\dif t} = -\left(H\frac{p^2}{E}+\dot{\Phi}\frac{p^2}{E}+\frac{p\hp^i}{a}\frac{\partial \Psi}{\partial x^i}-\me_0\frac{\hp^i}{a}\frac{p^2}{E}\frac{\partial \phi}{\partial x^i}\right). \label{dEdt}
}
In the equation above, the dot $\dot{\ }$ denotes the derivative with respect to cosmic time $t$, $H=\dot{a}/a$ represents the Hubble parameter and $\phi$ means electric potential whose gradient relates to
the electric field $\mathbf{E}=-\nabla \phi$ appeared in \Eq{electrodynamics2}. Instituting Eqs.~\eqref{4P}, \eqref{dxidt} and \eqref{dEdt} into \Eq{dfe}, there is
\eq{
    \frac{\partial f_\me}{\partial t}+\frac{\hp^i}{a}\frac{p}{E}\frac{\partial f_\me}{\partial x^i}-\frac{\partial f_\me}{\partial E}
    \left[ H\frac{p^2}{E}+\frac{\partial \Phi}{\partial t}\frac{p^2}{E}+
    \frac{p\hp^i}{a}\frac{\partial \phi}{\partial x^i}-\me_0\frac{1}{a}\frac{p^2\hp^i}{E}\frac{\partial \phi}{\partial x^i}\right]=0. \label{VE}
}
This is the Vlasov equation that describes a electron's statistical state in phase space forced by a electric field moving in a perturbed FRW Unierse (having neglected the collision term). This equation in flat spacetime has been widely and deeply studied
in the area of plasma physics and mathematics.

\section{\label{primordial}Damping of primordial electric field before decoupling}                                             

\subsection{Linearized Vlasov equation\label{linearizedVE}}
\Eq{VE} describes an electron's Boltzmann equation in electric field with respect of cosmic time $t$, but in this section the derivative to proper time $\lambda$ is quite convenience
on the calculations. Define the three-velocity with a new symbol
\eq{
    v^i=\frac{\dif x^i}{\dif \lambda}.
}
The complete differential of electron's state in phase space reads
\eq{
    \frac{\dif f_\me}{\dif \lambda}=\frac{\partial f_\me}{\partial \lambda}+\frac{\partial f_\me}{\partial \bx}\cdot\frac{\dif \bx}{\dif \lambda}
        +\frac{\partial f_\me}{\partial \bv}\cdot\frac{\dif \bv}{\dif \lambda}=C.     \label{dfedlambda}
}
For convenience and easy understanding, from now on in this section, the symbol of proper time $\lambda$ is replaced by the symbol $t$, and it means proper time instead of cosmic time.
Applying the geodesic equation \eqref{electrodynamics1} by setting the component $\mu=i$ and inserting it into the Boltamann equation \eqref{dfedlambda}, it arrives
\eq{
    \dot{f}_\me+\bv\cdot\frac{\partial f_\me}{\partial \bx}-\left[H\bv+\dot{\Phi}\bv+\frac{1}{a^2}\nabla\Psi+\frac{\me_0}{m}\bE\right]
        \cdot\frac{\partial f_\me}{\partial \bv}=C, \label{VEproper}
}
where the the dot~$~\dot{~}~$~in this section represents the derivative to proper time $\lambda$. In \Eq{VEproper}, the following equations or relations have
been used: $\Gamma^i_{\;00}=\partial_i \Psi/a^2$, $\Gamma^i_{\;j0}=\delta_{ij}(H+\partial\Phi/\partial t)$ and $\dif t/\dif \lambda\simeq1$ since nonrelativistic limit.
The collision term in \Eq{VEproper} is decomposed into two parts $C=-\varepsilon f_\me+n_\me\int\dif\Omega\,\partial f_{\me}/ \partial\Omega$.
The part including the angular integral in collision is independent on state function $f_\me$ and it is only a nonhomogeneous term, so in the following computation it is not taken into account.

The amplitude of electric field is so weak that the deviation to the equilibrium about electron's distribution function is quite slight:
\eq{
    f_\me(\bx,\bv,t)=\neo(\bv,t)+\delta \nee(\bx,\bv,t), \label{fe}
}
where $\neo$ is the probability density function under equilibrium state in absence of scaler perturbations in \Eq{metric}. instituting \Eq{fe} into \Eq{VEproper} and consider only the first order,
together with the Maxwell equation, it shows the equations
\eq{
    &\frac{\partial \delta n}{\partial t}+\bv\cdot\frac{\partial \delta n}{\partial \bx}-\left[H\bv+\dot{\Phi}\bv+\frac{1}{a^2}\nabla\Psi+\frac{\me_0}{m}\bE\right]
        \cdot\frac{\partial \neo}{\partial \bv}=-\varepsilon \delta n, \label{VE_p_1_order} \\
    &\quad\quad\quad\quad\rho(\bx,t)=-\me_0\int\delta n(\bx,\bv,t)\dif^3 v, \label{rho}\\
    &\quad\quad\quad\quad\mathbf{j}(\bx,t)=-\me_0\int\delta n(\bx,\bv,t)\bv\dif^3 v, \label{j}\\
    &\quad\quad\quad\quad\quad\quad\quad\nabla\cdot\bE=4\pi\rho(\bx,t),\\ \label{nablaE}
    &\quad\quad\quad\quad\frac{\partial\bE}{\partial t}=-4\pi\mathbf{j}(\bx,t),\ \ \ \ \nabla\times\bE=0.
}
$\rho$ in \Eq{rho} is the charge density while $\mathbf{j}$ in \Eq{j} is the current density.

Set the perturbed probability density function shows an oscillating form
\eq{
    \delta n=\delta n(\bv)\me^{\mi\left[\bk\cdot\bx-\int^t\omega(t')\dif t'\right]}. \label{delta_n}
}
Therefor, \Eq{VE_p_1_order} becomes by neglecting $\Phi$ and $\Psi$
\eq{
    -\mi\omega\delta n+\mi\bv\cdot\bk\delta n+\varepsilon \delta n=\left[H\bv+\frac{\me_0}{m}\bE\right]\cdot\frac{\partial n_0}{\partial \bv}. \label{delta_n_k}
}
Then, \Eq{delta_n_k} gives the fractional solution to perturbed probability density function:
\eq{
    \delta n=\frac{1}{\mi[\bk\cdot\bv-\omega(t)-\mi\varepsilon]}\left[H\bv+\frac{\me_0}{m}\bE\right]\cdot\frac{\partial n_0}{\partial \bv}. \label{delta_n_k_1}
}

\subsection{Polarization of plasma gas\label{PolarizationPG}}
Polarization vector indicates the deviation from electric neutrality, and it defines
\eq{
    \mathbf{D}_0 = \bE_0+4\pi\bP,
}
where $\mathbf{D}$ is the electric displacement vector which connects the electric field $\bE$ with permittivity tensor $\varepsilon_{ij}$:
\eq{
    D_{0i}\exp&\left[\mi\left(\bk\cdot\bx-\int^t\omega(t')\dif t'\right)\right]=\nonumber \\
         &\varepsilon_{ij}(\omega,\bk,t)E_{0j}\exp\left[\mi\left(\bk\cdot\bx-\int^t\omega(t')\dif t'\right)\right]. \label{electric_displacement}
}
The permittivity tensor could be decomposed into two components, the transverse one and longitudinal one. In this paper, the only component, transverse permittivity $\varepsilon_1$, is used which
means the direction of electric field $\bE_0$ is parallel to the direction of waves $\bk$, or $\bE_0\varparallel\bk$. So, polarization vector is
\eq{
    4\pi\bP_0=(\varepsilon_1-1)\bE_0. \label{polarization_vector}
}
According to \Eq{ne}, polarization vector is able to define in such a way:
\eq{
    \nabla\cdot\bP &= -a^3 n_\me-\int^t a^3\dot{\Phi}\dif t', \label{nablaP} \\
    \frac{\partial \bP}{\partial t} &= a^2n_\me \bv. \label{Pt}
}
Of course, the definitions above indicate $\bP$  propagates as a form of oscillating wave
\eq{
    \bP=\bP_0\exp\left[\mi\left(\bk\cdot\bx-\int^t\omega(t')\dif t'\right)\right].  \label{Pwave}
}
Thus, inserting \Eq{delta_n_k_1} and \eqref{Pwave} into \Eq{nablaP} gives
\eq{
    \mi\bk\cdot\bP&=\me_0 a^3\int \frac{\dif^3v}{\mi(\bk\cdot\bv-\omega)}\frac{\partial n_0}{\partial \bv}\cdot\left(H\bv+\frac{\me_0}{m}\bE\right) \nonumber \\
    &=\frac{\me_0^2}{m}a^3\bE\cdot\int\frac{\dif^3 v}{\mi(\bk\cdot\bv-\omega-\mi\varepsilon)}\frac{\partial n_0}{\partial \bv}
    +\me_0Ha^3\int\frac{\dif^3 v}{\mi(\bk\cdot\bv-\omega-\mi\varepsilon)}\frac{\partial n_0}{\partial \bv}\cdot\bv. \label{ikP}
}
The equation above is rewritten in a short form $\mi\bk\cdot\bP=\bE\cdot\mathbf{K}+\mathbf{K}_0$, compared which with \Eq{polarization_vector} gives the expression to the transverse permittivity tensor
\eq{
    \varepsilon_1(\omega,\bk,t)=1-\frac{4\pi\me_0^2}{mk^2}a^3\int\frac{\dif^3 v}{\mi(\bk\cdot\bv-\omega-\mi\varepsilon)}\frac{\partial n_0}{\partial \bv}\cdot\bk. \label{varepsilon1}
}
By applying \Eq{Pt}, it gives the expression of the longitudinal permittivity tensor
\eq{
    \varepsilon_t(\omega,\bk,t)=1+\frac{4\pi\me_0^2}{m\omega}a^3\int\frac{n_0(\bv)\dif^3 v}{\mi(\bk\cdot\bv-\omega-\mi\varepsilon)},
}
but this formula will not be used any longer, so there's no need to show any more interpretations.

\subsection{Expression of electric fields\label{Electric_fields}}

As discussed above, density function of electron could contain a slight perturbation as a function of $t$
\eq{
    n(\bx,\bk,t)=n_{0}(\textbf{v},t)+\delta n(\textbf{r},\textbf{v},t),
}
whose initial condition is
\eq{
    n(\bx,\bk,0)=n_{0}(\textbf{v},0)+g(\textbf{r},\textbf{v},0). \label{ninitial}
}
Start with \Eq{VE_p_1_order}, \eqref{rho} and \eqref{nablaE}, perturbed density function $\delta n$ and electric potential $\varphi$ satisfy the equations
\eq{
    &\frac{\partial \delta n}{\partial t}+\bv\cdot\frac{\partial \delta n}{\partial \bx}-\left[H\bv+\dot{\Phi}\bv+\frac{1}{a^2}\nabla\Psi+\frac{\me_0}{m}\bE\right]
        \cdot\frac{\partial \neo}{\partial \bv}=-\varepsilon\delta n, \label{deltan}\\
    &\nabla^2\varphi=4\pi\me_0\int\delta n(\bx,\bv,t)\dif^3 v. \label{varphi}
}
$\delta n$ and $\varphi$ are transformed into wave number vector space by Fourier transformation
\eq{
    \delta n_\bk(\bv,t)=\int_{-\infty}^{+\infty} \delta n(\bx,\bv,t)\me^{\mi\bk\cdot\bx}\dif^3 x,
}
and
\eq{
    \varphi_\bk(t)=\int_{-\infty}^{+\infty} \varphi(\bx,t)\me^{\mi\bk\cdot\bx}\dif^3 x.
}
Switching \Eq{deltan} and \eqref{varphi} into Fourier space by neglecting the perturbed scaler variables $\Phi$ and $\Psi$
\eq{
    \frac{\partial \delta n_\bk(\bv,t)}{\partial t}&+\mi\bk\cdot\bv\delta n_\bk(\bv,t)+\varepsilon\delta n_\bk(\bv,t)-
    \left[H\bv-\frac{\me_0}{m}\varphi_\bk(t)\mi\bk\right]\cdot\frac{\partial n_0(\bv)}{\partial\bv}=0 \label{delta_n_k_2}
}
and
\eq{
    a^{-2}k^2\varphi_\bk(t) = 4\pi\me_0\int\delta n_\bk(\bv,t)\dif^3 v \label{}
}
Then, Laplace transforms on $\delta n_\bk(\bv,t)$ and $\varphi_\bk(t)$ about proper time $t$ give
\eq{
    \delta_{\bk,s}(\bv)=\int_0^\infty \me^{-st}n_\bk(\bv,t)\dif t \label{deltanL}
}
and
\eq{
    \varphi_{\bk,s}=\int_0^\infty \me^{-st}\varphi_\bk(t)\dif t, \label{varphis}
}
whose inverse transformation is
\eq{
    \varphi_\bk(t)=\frac{1}{2\pi\mi}\int_{\sigma-\mi\infty}^{\sigma+\mi\infty} \varphi_{\bk,s}\me^{st}\dif s=\frac{1}{2\pi\mi}\sum_{i}\textrm{Res}\left( \varphi_{\bk,s_i}\me^{s_it}\right) \label{inverseVarphis}
}
with $\textrm{Res}(\cdots)$ denoting the $i$'th residual of integrand $\varphi_{\bk,s}\me^{st}$.
While the Laplace transform on the first term $\partial \delta n_\bk(\bv,t)/\partial t$ including a time derivative in \Eq{deltan} reads
\eq{
    \int_0^\infty \frac{\partial \delta n_\bk(\bv,t)}{\partial t}\me^{-st}\dif t=s\delta n_{\bk,s}(\bv)+g_\bk(\bv), \label{deltans}
}
where $g_\bk(\bv)$ represents the Fourier transformation on $g(\bx,\bv,0)$.

Multiply $\me^{-st}$ on both sides of \Eq{delta_n_k_2} and then integrate from $0$ to $\infty$. Using \Eq{deltans}, then \Eq{delta_n_k_2} becomes an equation as a function of parameter $s$:
\eq{
    s\delta n_{\textbf{k},s}(\textbf{v})-&g_{\textbf{k}}(\textbf{v})+\mi\textbf{k}\cdot\textbf{v}\delta n_{\textbf{k},s}
    +\varepsilon\delta n_{\textbf{k},s}(\textbf{v})-H\textbf{v}\cdot\frac{\partial n_0}{\partial \textbf{v}}\nonumber\\
    &+\frac{4\pi \me_0^{2}}{mk^{2}}\mi a^{3}\textbf{k}\cdot\frac{\partial n_{0}}{\partial \textbf{v}}\int\delta n_{\textbf{k},s}(\textbf{v}')\dif^{3}v'=0. \label{delta_n_s_1}
}
Collate the equation above, and integrate over all $\bv$ to find the solution
\eq{
    &\left(1-\frac{4\pi e_{0}^{2}a^{3}}{mk^{2}}\int\frac{1}{\textbf{k}\cdot\textbf{v}-is-i\varepsilon}\textbf{k}\cdot\frac{\partial n_{0}}{\partial
                \textbf{v}}\dif^{3}v\right)\int\delta n_{\textbf{k},s}(\textbf{v}')\dif^{3}v'\nonumber\\
    =&\int\frac{g_\textbf{k}(\textbf{v})}{s+\mi \textbf{k}\cdot\textbf{v}+\varepsilon}\dif^{3}v+
    H\int\frac{1}{s+\mi\textbf{k}\cdot\textbf{v}+\varepsilon}\textbf{v}\cdot\frac{\partial n_{0}}{\partial\textbf{v}}\dif^{3}v. \label{delta_n_s_2}
}
Multiplying $4\pi\me_0/k^2$ on both sides of \Eq{delta_n_s_2} and applying the definitions on transverse permittivity $\varepsilon_1(\omega,\bk,t)$ in \Eq{varepsilon1}
and electric potential $\varphi_{\bk,s}$ in \Eq{varphis}, the electric potential is obtained
\eq{
    \varphi_{\bk,s}=\frac{\mi 4\pi \me_{0}a^{2}}{k^{2}\varepsilon_{1}(\mi s,\textbf{k})}\left(\int\frac{g_\textbf{k}(\textbf{v})\dif^{3}v}{\textbf{k}\cdot\textbf{v}-\mi s-\mi\varepsilon}+\int\frac{H\textbf{v}
    \cdot\frac{\partial n_{0}}{\partial\textbf{v}}\dif^{3}v}{\textbf{k}\cdot\textbf{v}-\mi s-\mi\varepsilon}\right). \label{delta_n_s_3}
}

It notes the Laplace transformations above play only on the oscillating variables, like $\delta n$, but avoid the scalar factor $a$. This is because the intrinsic oscillating
frequency $\Omega_\me$ is extremely larger than the Hubble expansion rate $H$ as discussed at the of this section.
Besides, the exactly analytic expression of $a$ is not going to be involved at all in this section.

\subsection{Plasma with Maxwell distribution\label{Maxwell}}

The electron's mass is much lighter than the background particles, like proton, so it's reasonable to assume electrons is on the equilibrium which follow the Maxwell distribution \cite{cercignani2012relativistic}
\eq{
    n_{\me}(\bv)= {N_{\me}}{\sqrt{\mathrm{det} g_{ij}}}\left(\frac{m}{2\pi \kB T}\right)^{\frac{3}{2}}\exp\left(-\frac{g_{ij}P^iv^j}{2\kB T}\right),
}
here $g_{ij}=a^2\delta_{ij}$ denoting the background metric and $N_{\me}\propto a^{-3}$ denoting the particle number density.
Define a new variable about velocity $\tilde{\bv}=a\bv$, then the distribution functions follows
\eq{
    \tn_0(\tilde{\bv})=N_{\me0}\left(\frac{m}{2\pi \kB T}\right)^{\frac{3}{2}}\exp\left(-\frac{m\tilde{\bv}^2}{2\kB T}\right), \label{ne0}
}
with $N_{\me0}\equiv N_{\me}a^3$ representing electron's number density at present time. Integrating \Eq{ne0} over $\tv_y$ and $\tv_z$ generates the Maxwell distribution along $x$ direction
\eq{
    \tn_0(\tv_x)= N_{\me0}\left(\frac{m}{2\pi \kB T}\right)^{\frac{1}{2}}\exp\left(-\frac{m\tv_x^{2}}{2\kB T}\right). \label{ne0x}
}

With the discussions above and setting $\bk$ parallel to $x$-axis, the transverse permittivity could be reexpressed as follow:
\eq{
    \varepsilon_1=1-\frac{4\pi e_0^2a}{k^2m}\int k\frac{\partial \tn_0}{\partial \tv_x}\frac{\dif\tv_x}{k_\rp\tv_x-\omega-\mi\varepsilon}, \label{varepsilon1_2}
}
with definition of physical wave number $k_\rp=k/a$. Generally, the ratio $\varepsilon/k_\rp\ll1$, so it is labeled as $0^+$ below for convenience. Instituting \Eq{ne0x} into \Eq{varepsilon1_2}, it arrives
\eq{
    \varepsilon_1=1-&\frac{4\pi e_{0}^{2}}{k_\rp^2 m}\int \frac{\dif \tv_x}{\tv_x-\frac{\omega}{k_\rp}-\mi0}
     N_{\me0}\left(\frac{m}{2\pi \kB T}\right)^{\frac{1}{2}}\left(-\frac{m\tv_x}{\kB T}\right)\exp\left(-\frac{m\tv_{x}^{2}}{2\kB T}\right). \label{varepsilon1_3}
}
It's convenient to define the following parameters for the subsequent computations
\eq{
\begin{cases}
    \lambda_\me^{-1}=\sqrt{\dfrac{4\pi e_{0}^{2} N_{\me0}}{\kB T}},\quad v_\me=\sqrt{\dfrac{\kB T}{m}}, \\
    z=\dfrac{\tv_x}{\sqrt{2}v_\me},\quad\quad x=\dfrac{\omega}{\sqrt{2}k_\rp v_\me}. \label{parameters}
\end{cases}
}
The parameters in the equations above used to reflect the plasma characteristics are $\lambda_\me$, which means the Debye radius at the moment of recombination,
and $v_\me$, which corresponds to the average thermal velocity.
Thus \Eq{varepsilon1_3} reexpresses
\eq{
    \varepsilon_1&=1+\frac{1}{\kp^2\lambda_\me^2}\int\frac{\dif\tv_x}{\tv_x-\frac{\omega}{\kp}-\mi0} \frac{1}{\sqrt{2\pi}} \frac{\tv_x}{v_\me} \exp\left(-\frac{\tv_x^2}{2v_\me^2}\right)\nonumber\\
                 &=1+\frac{1}{\kp^2\lambda_\me^2}\frac{1}{\sqrt{\pi}}\int^{\infty}_{0}\frac{z}{z-x-\mi0}e^{-z^{2}}\dif z \nonumber\\
                 &=1+\frac{1}{\kp^2\lambda_\me^2}\frac{1}{\sqrt{\pi}}\int^{\infty}_{0}\frac{(z-x)+x}{z-x-\mi0}e^{-z^{2}}\dif z \nonumber\\
                 &=1+\frac{1}{\kp^{2}\lambda_\me^2}\left[1+\frac{x}{\sqrt{\pi}}\int^{\infty}_{0}\frac{\me^{-z^{2}}}{z-x-\mi0}\dif z\right]\nonumber\\
                 &=1+\frac{1}{\kp^{2}\lambda_\me^2}[1+F(x)]. \label{varepsilon1_4}
}
In the last equality, integral of Gaussian type \eqref{Gauss1} has been used and $F(x)$ is a function as a form of Cauchy's type
\eq{
    F(x)\equiv\frac{x}{\sqrt{\pi}}\int^{\infty}_{0}\frac{\me^{-z^{2}}}{z-x-\mi0}\dif z. \label{Fx}
}
According to the introductions of Cauchy's type in Appendix \ref{Cauchy}, the permittivity at high frequency limit $\omega\gg \kp v_\me$,
or $x\gg1$, approximates
\eq{
    \varepsilon_{1}\approx1-\frac{\Omega_\me^2}{\omega^{2}}\left(1+\frac{3\kp^{2}v_\me^2}{\omega^{2}}\right)+\mi\sqrt{\frac{\pi}{2}}
        \frac{\omega\Omega_{\me}^{2}}{(\kp v_{\me})^{3}}\exp\left(-\frac{\omega^{2}}{2\kp^{2}v_{\me}^{2}}\right),  \label{varepsilon1_high}
}
where
\eq{
    \Omega_\me\equiv\frac{v_\me}{\lambda_\me}=\sqrt{\frac{4\pi N_{\me0}e_{0}^{2}}{m}}
}
is defined as the Langmuir frequency. In the low frequency limit $x\ll 1$, it evaluates
\eq{
    \varepsilon_1\approx1+\left(\frac{\Omega_\me}{\kp v_\me}\right)^2\left[1-\left(\frac{\omega}{\kp v_\me}\right)^2
        -\mi\sqrt{\frac{\pi}{2}}\frac{\omega}{\kp v_\me}\right],\ \ (x\ll1).\label{varepsilon1_low}
}
Applying parameters in \Eq{parameters}, the second part of electric potential in \Eq{delta_n_s_3} arrives
\eq{
    \varphi_{\bk,s}^{(2)}=&\frac{H}{\kp}\frac{\mi4\pi e_{0}}{k^2a\varepsilon_1}\int\frac{\dif\tv}{\tv-\frac{\mi s}{\kp}-\mi0}\cdot\frac{\partial\tn_0(\tv_x)}{\partial\tvx}\nonumber\\
                          &+2\frac{H}{\kp}\frac{\mi4\pi e_{0}}{k^{2}a\varepsilon_{1}}\int\frac{\dif^2\tv}{\tvx-\frac{\mi s}{\kp}-\mi0}N_{\me0}\left(\frac{m}{2\pi \kB T}\right)^{\frac{3}{2}}\nonumber\\
                          &\ \ \ \ \ \times\exp\left(-\frac{m(\tvx^2+\tv_y^2+\tv_z^2)}{2\kB T}\right)\cdot\left(-\frac{m\tv_y^2}{\kB T}\right)\nonumber\\
                         =&\frac{H}{\kp}\frac{v_\me m}{\mi a^3 e_0\varepsilon_{1}}\frac{4\pi e_0^2 N_{\me0}}{\kp^{2}\kB T}
                                        \left[\sqrt{\frac{2}{\pi}}\int_0^{+\infty}\frac{z^2\me^{-z^2}}{z-x-\mi0}\dif z\right. \nonumber\\
                          &\left.\quad + 2\int_0^{+\infty}\frac{\me^{-z^2}}{z-x-\mi0}\frac{\dif z}{\sqrt{2\pi}}\cdot\int_0^{+\infty}z'^2\me^{-z'^2}\dif z'\right]\nonumber\\
                         =&\frac{1}{\mi\varepsilon_1(\omega,\bk)}\frac{\sqrt{2}H v_\me m}{\kp a^{3}e_0}\frac{1}{\kp^{2}\lambda_\me^2}
                                        \left[\frac{1}{\sqrt{\pi}}\int_0^{+\infty}\frac{z^{2}\me^{-z^{2}}}{z-x-\mi0}\dif z\right. \nonumber\\
                          &\left.\quad\quad\quad\quad\quad\quad\quad\quad   +\frac{1}{\sqrt{\pi}}\int\frac{\me^{-z^2}}{z-x-\mi0}\dif z\right]\nonumber\\
                         =& \frac{1}{\mi\varepsilon_1(\omega,\bk)}\frac{1}{\kp^3\lambda_\me^2\lambda_H}\left[1+xF(x)+2F(x)+\frac{F(x)}{x}\right], \label{varphi(2)}
}
where
\eq{
    \lambda_H^{-1} \equiv \frac{\sqrt{2}H v_\me m}{\kp a^3e_0}
}
is a parameter relates to Hubble expansion rate, and \Eq{Gauss1} has been used during the calculations above.

Finally, we get the exact expression of electric potential under FRW metric in absence of $\Phi$ and $\Psi$
\eq{
    \varphi_{\bk,s}=\frac{\mi4\pi e_0}{\kp^2}\frac{\chi(\mi s,\bk)}{\varepsilon_1(\mi s,\bk)}
        -\frac{\mi}{\kp^3\lambda_\me^2\lambda_H}\frac{G({\mi s}/{\sqrt{2}v_\me \kp})}{\varepsilon_1(\mi s,\bk)}, \label{varphi_final}
}
where
\eq{
    \chi(\mi s,\bk)=\int\frac{\tilde{g}_\bk(v_x)\dif v_x}{kv_x-\mi s-\mi0}
}
and
\eq{
    G(x)=1+xF(x)+2F(x)+{F(x)}/{x}. \label{Gx}
}
\subsection{Diffusion damping of primordial electric fields\label{Landau}}

During the epoch before decoupling, electrons state on a thermal equilibrium with photons, but the protons have already exited the thermal equilibrium state with photons for along period,
so $T_\me\gg T_\textrm{p}$. Thus the high frequency limit means $\omega\gg\kB T_\me\gg\kB T_\textrm{p}$. \Eq{varepsilon1_high} rewrites by setting $\beta\equiv \kp v_\me/\omega\ll1$
\eq{
    \frac{\omega^2}{\Omega_\me^2}=1+3\beta^2-\mi\sqrt{\frac{2}{\pi}}\beta^{-3}\me^{-1/2\beta^2}. \label{varepsilon1_high_beta}
}
The first iteration gives the first order zero point
\eq{
    \omega=\Omega_\me
}
Then the second time gives the solution where the term $\beta^2$ has taken into account:
\eq{
    \omega=\Omega_\me(1+3\kp^2\lambda_\me^2)^{1/2}. \label{second}
}
Finally, inserting \Eq{second} into \Eq{varepsilon1_high_beta}, we get a complex solution $\omega=\omega'+\mi\omega''$, with
\eq{
    \omega'=\Omega_\me\left(1+\frac{3}{2}\kp^2\lambda_\me^2\right), \label{omega'}
}
\eq{
    \omega''=-\sqrt{\frac{\pi}{8}}\frac{\Omega_\me}{(\kp\lambda_\me)^3}\exp\left[-\frac{1}{2(\kp\lambda_\me)^2}-\frac{3}{2}\right]. \label{omega''}
}
Similarly, the zero point at low frequency is
\eq{
    \omega'=\kp v_\me,\ \ \text{and}\
    \omega''=-\sqrt{\frac{\pi}{8}}\omega'.
}
Thus the root of equation $\varphi_{\mi s,\bk}=0$ reads $s=\mi\omega'-\gamma$ with $\gamma=-\omega''$. Base on \Eq{inverseVarphis},  it leads to the expression of electric potential
\eq{
    \varphi_{\bk}(t)=\frac{2e_0}{\kp^2}\sum_i R_i\me^{s_it}-\frac{1}{2\pi\kp^3\lambda_\me^2\lambda_H}\sum_i R'_i\me^{s_it}. \label{varphi_si}
}

Obviously, the potential is going to damp out by Landau damping, so the initial condition before electron's recombination, like neutrino's decoupling,
affects almost nothing on cosmic microwave background. Although the potential may be a large amplitude on large scale, it still damps away as well with time increasing.

On the other hand, although the electric potential damps out with time increasing, the particle density $\delta n$ has an oscillating mode.
Inserting \Eq{delta_n_s_3} into \Eq{delta_n_s_1}, we have
\eq{
    \delta_{\bk,s}=\frac{1}{s+\mi\bk\cdot\bv}&\left\{g_\bk(v)+\left[H\bv+\frac{4\pi e_0^2}{mk^2}
    \left(\frac{\chi(\mi s,\bk)}{\varepsilon_1(\mi s,\bk)}
        +\frac{\varsigma(\mi s,\bk)}{\varepsilon_1(\mi s,\bk)}+\frac{C(\mi s,\bk)}{\varepsilon_1(\mi s,\bk)} \right)\bk \right]\cdot\frac{\partial n_0}{\partial \bv}\right\}, \label{delta_n_2}
}
where $C(\mi s,\bk)$ is the Laplace and Fourier transform on the nonhomogeneous part in collision term. Except the singularities at $\varepsilon=0$,
the point $s=-\mi\bk\cdot\bv$ is also a singularity. So the inverse Laplace transformation on \Eq{delta_n_2} gives an oscillating mode without damping
\eq{
    \delta n_\bk(\bv,t)\propto\me^{-\mi\bk\cdot\bv}.
}

Furthermore, it's necessary to characterize the attenuating scale of electric fields by estimating the order of $a^{3/2}\lambda_\me H$ or $a^{-3/2}\Omega_\me/H$:
\eq{
    \frac{H^2}{(\Omega_\me a^{-3/2})^2}=\frac{\varepsilon_0 mc^2 a^3}{4\pi e_0^2N_{\me0}}\frac{H^2_0}{c^2}\frac{H^2}{H_0^2}.\label{lambdaH}
}
This number is covenient to calculate under natural unit system: $\varepsilon_0/4\pi e_0^2=2.22\times10^{45}\;\mathrm{kg}^{-1}\cdot\mathrm{m}^{-3}\cdot\mathrm{s}^{2}$,
$mc^2=0.51\;\mathrm{MeV}=8.16\times10^{-14}\;\mathrm{kg}\cdot\mathrm{m}^{2}\cdot\mathrm{s}^{-2}$, $N_{\me0}=11.244X_\me\Omega_\text{m}h^2\;\mathrm{m}^{-3}$ and
$H_0/c=3.33\times10^{-4}h\;\text{Mpc}^{-1}$. At early epoch of the Universe, the main component is either radiation or matter, so $H^2/H_0^2=\Omega_\text{m}a^{-3}[1+a_\text{eq}/a]$.
Then \Eq{lambdaH} becomes
\eq{
    \frac{a^3H^2}{\Omega_\me^2}=2.11\times10^{-20}\frac{\Omega_\text{m}h^2}{X_\me\Omega_\text{b}h^2}\left(1+\frac{a_\text{eq}}{a}\right).\label{lambdaH1}
}
Recent Planck data shows $\Omega_\text{m}h^2=0.1424$ and $\Omega_\text{b}h^2=0.022447$ \cite{refId0,refId1}. Saha equation predicts $X_\me\sim10^{-2}$ at present time.
With the discussions above, \Eq{lambdaH} estimates
\eq{
    {a^3H^2}/{\Omega_\me^2}=1.91\times10^{-19}\left(1+\frac{a_\text{eq}}{a}\right)\sim10^{-18},
}
thus ${a^{3/2}H}/{\Omega_\me}\sim10^{-9}$. While, it also estimates $a^{3/2}\lambda_\me H\sim1$. This estimation indicates that the electric field entering Hubble horizon damps out immediately by Landau damping effect.
So the primordial electric fields give not any effects on the power spectra of cosmic microwave background.

\section{Soliton of electric fields inside plasma before decoupling\label{soliton}}

In this section, let's discuss the nonlinear effect in the primordial plasma.
\subsection{Basic equations via moment method}
Back to the Vlasov equation appeared in \Eq{VE} in terms of derivative to cosmic time $t$ instead of proper time in Sec.~\ref{primordial}.
Define two variables
\eq{
    n_\me\equiv\int\frac{\dif^3p}{(2\pi)^3}f_\me,\label{ne2}
}
and
\eq{
    n_\me v^i\equiv\int\frac{\dif^3p}{(2\pi)^3}\frac{p\hp^i}{E}f_\me.\label{nev}
}
Multiply $\dif^3p/(2\pi)^3$ on both sides of \Eq{VE} and integrate over all momentum space,
\eq{
    &\quad\frac{\partial}{\partial t} \int \frac{\dif^{3} p}{(2 \pi)^{3}} f_\me  +  \frac{1}{a} \frac{\partial}{\partial x^{i}} \int \frac{d^{3} p}{(2 \pi)^{3}} \frac{p \hat{p}^{i}}{E}f_\me  \nonumber\\
    &-\left[H+\dot{\Phi}\right] \int \frac{\dif^{3} p}{(2 \pi)^{3}} \frac{\partial f_\me}{\partial E} \frac{p^{2}}{E} -\frac{1}{a} \frac{\partial \Psi}{\partial x^i}\int \frac{\dif^{3} p}{(2 \pi)^{3}} \frac{\partial f_{\mathrm{dm}}}{\partial E} \hat{p}^{i} p \nonumber \\
    &+\frac{e_0}{a}\partial_i\phi \int \frac{\dif^{3} p}{(2 \pi)^{3}}\frac{p \hat{p}^{i}}{E}f_\me=0, \label{eq0}
}
The third term on the left hand integrates by part
\eq{
    \int\frac{\dif^3p}{(2\pi)^3}\frac{p^2}{E}\frac{\partial f_\me}{\partial E} &= \int\frac{\dif^3p}{(2\pi)^3}\frac{\partial f_\me}{\partial p}p\nonumber\\
    &={-3}\frac{4\pi}{(2\pi)^3}\int_{0}^{+\infty}p^2f_\me\dif p\nonumber\\&={-3}n_\me, \label{eq1}
}
where the relation $(p/E)(\partial f_\me/\partial E)=\partial f_\me/\partial p$ has been used in first equality.
Because electrons during the matter dominated epoch are nonrelativistic particles, $v^i$ is regarded as a first order variable.
Inserting Eqs.~\eqref{ne2}, \eqref{nev} and \eqref{eq1} into \Eq{eq0} and neglecting the terms higher than first order, it reduces to
\eq{
    \frac{\partial n_\me}{\partial t}+3[H+\dot{\Phi}]n_\me+\frac{1}{a}\partial_i(n_\me v^i)=0.  \label{ne}
}
This is called the zero order moment equation.

Multiplying $\dfrac{\dif^3p}{(2\pi)^3}\dfrac{p\hp^i}{E}$ on both sides of \Eq{VE} and integrating over $\bp$, this gives the first order moment equation
\eq{
    &\quad\quad\frac{\partial }{\partial t}\int\frac{\dif^3p}{(2\pi)^3}\frac{p\hp^j}{E}f_\me+\frac{1}{a}\frac{\partial}{\partial x^i} \int\frac{\dif^3 p}{(2\pi)^3}\frac{p^2\hp^i\hp^j}{E^2}f_\me\nonumber\\
    &-[H+\dot{\Phi}]\int\frac{\dif^3p}{(2\pi)^3}\frac{\partial f_\me}{\partial E}\frac{p^3\hp^j}{E}
        -\frac{1}{a}\frac{\partial\Psi}{\partial x_i}\int\frac{\dif^3p}{(2\pi)^3}\frac{\partial f_\me}{\partial E}\frac{\hp^i\hp^jp^2}{E}\nonumber\\
    &\quad\quad\quad+e_0\frac{1}{a}\frac{\partial\phi}{\partial x_i}\int\frac{\dif^3p}{(2\pi)^3}\frac{\partial f_\me}{\partial E}\frac{p^2\hp^i\hp^j}{E^2}=0. \label{eq2}
}
The second term in \Eq{eq2} reexpresses as
\eq{
    \int\frac{\dif^3}{(2\pi)^3}\frac{p^2\hp^i\hp^j}{E^2}f_\me = &\int\frac{\dif^3p}{(2\pi)^3} f_\me\left(\frac{p\hp^i}{E}-v^i\right)\left(\frac{p\hp^j}{E}-v^j\right)\nonumber\\
    &+\int\frac{\dif^3p}{(2\pi)^{3}}f_\me v^i v^j=P^{ij}+n_\me v^i v^j,\label{term2}
}
where $P^{ij}$ is called pressure tensor which corresponds to the anisotropy. The third term equals approximately to $-4\partial_i(n_\me v^j)/a$
by making use of the relation $(p/E)(\partial f_\me/\partial E)=\partial f_\me/\partial p$ and the fourth term equals to $n_\me\partial_j\Psi/a$.
The last integration is calculated by part explicitly as previous integrations. This integral is
\eq{
    \int\frac{\dif^3}{(2\pi)^3}\frac{p^2\hp^i\hp^j}{E^2}\frac{\partial f_\me}{\partial E}&=\int \dif\Omega \hp^i\hp^j\int_{0}^{\infty}\frac{\partial f_\me}{\partial p}\frac{p}{E}\nonumber\\
    &=-\int \dif\Omega \hp^i\hp^j\int_{0}^{\infty}\dif p\left(\frac{3p^{2}}{E}-\frac{p4}{E^{3}}\right)\nonumber\\&=-3\delta^{ij}\frac{n_\me}{m}, \label{term5}
}
where the additional fact is applied
\eq{
    \int \dif\Omega \hp^i\hp^j=\delta^{ij}\frac{4\pi}{3}.
}
Therefore, with the discussions above, \Eq{eq2} becomes
\eq{
    \frac{\partial(n_\me v^j)}{\partial t}+\frac{1}{a}&\partial_i(n_{e}v^iv^j)+\frac{1}{a}\partial_iP^{ij}+4Hn_\me v^j\nonumber\\
    &+\frac{n_\me}{a}\frac{\partial\Psi}{\partial x^j}-\frac{1}{a}\frac{e_{0}}{m}\frac{\partial\phi}{\partial x^j}=0. \label{1moment1}
}
Regroup the first two terms in the equation above
\eq{
    \quad\frac{\partial(n_\me v^j)}{\partial t}+\frac{1}{a}\frac{\partial(n_\me v^i v^j)}{\partial x^i}
    =\left[\frac{\partial n_\me}{\partial t}+\frac{1}{a}\frac{\partial (n_\me v^i)}{\partial t}\right]v^j+n_\me\left[\frac{\partial v^j}{\partial t}+\frac{1}{a}v^i\partial_i v^j\right]. \label{term12}
}
Instituting \Eq{ne} and \eqref{term12} into \Eq{1moment1} and neglecting the terms higher than first order but retaining the term $v^j\partial_j v^i$, it obtains the evolutionary equation of velocity
\eq{
    \dot{v}^i+Hv^i+\frac{1}{a}v^{j}\partial_{j}v^i+\frac{1}{a}\partial_{i}\Psi-\frac{1}{a}\frac{e_0}{m}\partial_{i}\phi=0. \label{vi}
}
Note that the terms containing $v^j\partial_j v^i$ is retained since the following computations will make use of it. \Eq{vi} is also able to describe a proton's velocity equation
just by exchanging $-e_0/m$ to $e_0/M$, with, of course, $M$ denoting the mass of photon.

\subsection{Nonlinear equations for electric fields\label{nonlinear}}

Electrons stay on a equilibrium state with photons and they move evidently much faster than protons, so electrons obey the Maxwell distribution
\eq{
    N_\me=N_{\me0}\exp\left(\frac{e_0\phi}{\kB T_\me}\right)\label{Ne}
}
with
\eq{
    N_{\me0}=\int n_\me(\bx,\bv,t)\dif^3v.
}
Thus we have the following equations
\begin{subequations}\label{solitonx}
\begin{numcases}{}
    \nabla^{2}\phi=4\pi e_0\left[N_{e0}\exp\left(\dfrac{e_0\phi}{\kB\Te}\right)-N_\rp\right],\label{soliton1_1}\\
    \dfrac{\partial N_\rp}{\partial t}+\dfrac{1}{a}\dfrac{\partial(N_\rp \vp^i)}{\partial x^i}+3[H+\dot{\Phi}]N_\rp=0,\label{soliton1_2}\\
    \dfrac{\partial v_\rp^2}{\partial t}+Hv_\rp^{i}+\dfrac{1}{a}\partial_{i}\Psi+v_\rp^{i}\partial_{i}v_\rp^{j}=-\frac{1}{a}\frac{e_{0}}{M}\partial_{i}\phi. \label{soliton1_3}
\end{numcases}
\end{subequations}
To nondimensionalize the equations, it's necessary to define such parameters as below:
\eq{
    &\tilde{\phi}=\frac{e_{0}\phi}{\kB\Te},\  \tN_{\me0}=a^3N_{\me0},\ \tN_\rp=a^{3}N_{p},\ \tT=aT,\ \tilde{N}=\frac{\tN_\rp}{\tN_\me}, \nonumber\\
    &\ \ \tlambdae^{-1}=\sqrt{\frac{4\pi e_0^2\tN_{\me0}}{\kB\tilde{T}_\me}},\ \tilde{x}=\frac{x}{\tlambdae},\ \tOmegai=\sqrt{\frac{4\pi e_0^2\tN_{\me0}}{M}},\ \ttt=\tOmegai t,\nonumber\\
    &\quad\quad\quad\quad\tu=\frac{v_\rp}{\sqrt{\kB\tT_\me/M}}\ \ \text{and}\ \ \dif\ttauu=\dif \tilde{t}/a. \label{parameters2}
}
Multiply $a^2/\kB\Te$, \Eq{soliton1_1} reduces to
\eq{
    \frac{\partial^2}{\partial x^{2}}\frac{e_0\phi}{\kB\Te}=\frac{4\pi e_0^2\tN_{\me0}}{\kB\tT_\me}\left[\exp\left(\frac{e_0\phi}{\kB T_\me}\right)-\frac{\tN_\rp}{\tN_{\me0}}\right].\label{eq3}
}
Inserting relevant parameters in \Eq{parameters2}, it leads to a nondimensional equation
\eq{
    \frac{\partial^2\tilde{\phi}}{\partial \tx^{2}}=e^{\tilde{\phi}}-\tilde{N}. \label{soliton2_1}
}
Multiply $a\tlambdae/(\tN_{\me0}\sqrt{\kB\tT_\me})$ on both sides of \Eq{soliton1_2}:
\eq{
    \frac{\tlambdae a}{\sqrt{\kB\tT_\me/M}}\cdot\frac{\partial}{\partial t}\left(\frac{\tN_\rp}{\tN_{\me0}}\right)+
    \frac{\tlambdae}{a}\frac{\partial}{\partial x}\left(\frac{\tN_\rp}{\tN_{\me0}} \cdot \frac{\tv_\rp}{\sqrt{{\kB\tT_\me}/{M}}}\right)=0,
}
which leads to
\eq{
    \frac{\partial\tN}{\partial\ttt}+\frac{1}{a}\frac{\partial}{\partial\tx}(\tN\tu)=0. \label{soliton2_2}
}
Similarly, \Eq{soliton1_3} becomes
\eq{
    \frac{\partial\tu}{\partial\ttauu}+\frac{\tu}{a}\frac{\partial\tu}{\partial\tx}=-\frac{\partial\tilde{\phi}}{\partial\tilde{x}}.\label{soliton2_3}
}

\subsection{Nonlinear solution\label{solution}}

The electrical neutrality condition provides $\tN=1+\tn$ with $\tn\ll1$. The linearized equations to \Eq{soliton2_1}, \eqref{soliton2_2} and \eqref{soliton2_3} read
\begin{subequations}
\begin{numcases}{}
    \frac{\partial \tilde{\phi}^2}{\partial \tilde{x}^2}=\tilde{\phi}-\tilde{n},\label{soliton3_1}\\
    \frac{\partial \tilde{n}}{\partial \tilde{\tau}}+\frac{1}{a}\frac{\partial \tilde{u}}{\partial \tilde{x}}=0,\label{soliton3_2}\\
    \frac{\partial \tilde{u}}{\partial \tilde{\tau}}+\frac{\partial \tilde{\phi}}{\partial \tilde{x}}=0. \label{soliton3_3}
\end{numcases}
\end{subequations}
Take derivative of \Eq{soliton3_1} to $\ttauu$ and insert \Eq{soliton3_2} into it,
\eq{
     \frac{\partial^3 \tilde{\phi}}{\partial \tilde{x}^2\partial\tilde{\tau}}
    =\frac{\partial \tilde{\phi}}{\partial \tilde{\tau}}-\frac{\partial \tilde{n}}{\partial \tilde{\tau}}
    =\frac{\partial \tilde{\phi}}{\partial \tilde{\tau}}+\frac{1}{a}\frac{\partial \tilde{u}}{\partial \tilde{x}}.\label{eq4}
}
Another derivative of \Eq{eq4} to $\ttauu$ gives
\eq{
    \frac{\partial^4 \tilde{\phi}}{\partial \tilde{x}^2\partial\tilde{\tau}^2}
    &=\frac{\partial^2\tilde{\phi}}{\partial\tilde{\tau}^2}+\frac{\partial^2\tilde{u}}{\partial\tilde{x}\partial\tilde{\tau}} \nonumber\\
    &= \frac{\partial^2\tilde{\phi}}{\partial\tilde{\tau}^2 }-\frac{\partial^2\tilde{\phi}}{\partial \tilde{x}^2}
    -\frac{\dif a^{-1}}{\dif \tilde{\tau}}\frac{\partial \tilde{u}}{\partial \tilde{x}}.
}
Since the Langmuir frequency $\tOmegai$ is much lager than the Hubble expansion rate $H$, i. e. $\tOmegai\gg H$, as discussed in Sec.~\ref{Landau}
\eq{
    \frac{\dif a^{-1}}{\dif \tilde{\tau}}=-\frac{1}{a^2}\frac{\dif a}{\dif t}\frac{\dif t}{\dif \tilde{\tau}}
    =- \frac{H}{\tOmegai}\ll1.
}
Thus we finally obtain the nonlinear partial differential equation
\eq{
    \frac{\partial \tilde{\phi}^2}{\partial \tilde{\tau}^2}-\frac{1}{a}\frac{\partial \tilde{\phi}^2}{\partial \tilde{x}^2}
     = \frac{\partial^4 \tilde{\phi}}{\partial \tilde{x}^2\partial\tilde{\tau}^2}. \label{solitary_wave_equation}
}
This equation describes a solitary wave meaning its profile would not change as it propagates. Setting
\eq{
    \tilde{\phi}\propto \me^{\mi\left[\tilde{k}\tilde{x}-\int^{\ttauu}\tilde{\omega}(\ttauu)\dif \ttauu\right]}
}
with $\tilde{k}=k\tlambdae$ and $\tilde{\omega}=\omega/\tOmegai$, inserting which into \Eq{solitary_wave_equation}, we get the dispersion relation
\eq{
    \omega^2=\frac{\tOmegai^2}{a}\frac{k^2\tlambdae^2}{1+k^2\tlambdae^2}
}
by applying the limit condition $H\ll(\dif\omega/\dif t)/\omega\ll\tOmegai$.
If consider only the linear part of \Eq{solitary_wave_equation} and return to natural unit, we have
\eq{
    \frac{\partial^2\tilde{\phi}}{\partial \tau^{2}}-\frac{\kB\tilde{T}_\me}{aM}\frac{\partial^{2}\tphi}{\partial x^{2}}=0.\label{linearsoliton}
}
Define the average thermal velocity
\eq{
    c_\textrm{s}=\sqrt{\kB T_\me/M}, \label{ecs}
}
which also stands for the phase velocity $c_\textrm{s}=a^{-1/2}\tlambdae\tOmegai$. $c_\textrm{s}$ is nether the thermal velocity of electrons nor protons, instead it's their
collective group velocity since the tight coupling with each other before recombination.
So solitary equation \eqref{solitary_wave_equation} propagates with velocity $c_\textrm{s}$ approximately.

Now give further analysis on \Eq{solitary_wave_equation}. Introduce the Mach number which represents the ratio of actual velocity to phase velocity
\eq{
    \rM\equiv\frac{\dif x/\dif\tau}{a^{1/2}c_\textrm{s}}=\frac{\dif x/\dif\tau}{\tlambdae\tOmegai}\propto a^{-\frac{1}{2}} .\label{M}
}
$\tphi$, $\tN$ and $\tu$ only dependent on a combination variable
\eq{
    \teta=\tx-\int^{\ttauu}\rM\dif\ttauu'.\label{eta}
}
Then equations in \eqref{solitonx} becomes a new set of dynamic equations in terms of $\eta$:
\begin{subequations}\label{soliton4}
\begin{numcases}{}
    \frac{\dif^2\tilde{\phi}}{\dif\teta^2}=\me^{\tilde{\phi}}-\tilde{N},\label{soliton4_1}\\
    -\rM\frac{\dif\tN}{\dif\teta}+\frac{1}{a}\frac{\dif(\tN\tu)}{\dif\teta}=0,\label{soliton4_2}\\
    -\rM\frac{\dif\tu}{\dif\teta}+\frac{\tu}{a}\frac{\dif\tilde{u}}{\dif\teta}=-\frac{\dif\tilde{\phi}}{\dif\teta},\label{soliton4_3}
\end{numcases}
\end{subequations}
with boundary condition $\tphi\to0$, $\dfrac{\dif\tphi}{\dif\teta}\to0$, $\tu\to0$, $\tN\to1$ and $\dfrac{\dif\tN}{\dif\teta}\to0$ when $|x|\to\infty$, $|\teta|\to\infty$.

Integrate \Eq{soliton4_2} over $\teta$ from $\teta$ to $\infty$
\eq{
    -\rM\int_{\eta}^{+\infty}\frac{\dif\tN}{\dif\teta'}\dif\teta'=\frac{1}{a}\int_{\teta}^{+\infty}\frac{\dif(\tN\tu)}{\dif\teta'}\dif\teta'
}
to get the first equation between $\tn$ and $\tN$
\eq{
    \tu=\rM a(1-\frac{1}{\tN}). \label{eq5}
}
Integrating on \Eq{soliton4_3}, we have
\eq{
    \left(\rM a^{\frac{1}{2}}-a^{-\frac{1}{2}}\tu\right)^{2}=\rM^{2}a-2\tilde{\phi}.
}
Instituting \Eq{eq5} into the equation above, it arrives
\eq{
    \tilde{N}=\frac{\rM a^{\frac{1}{2}}}{\sqrt{(\rM a^{\frac{1}{2}})^{2}-2\tilde{\phi}}}.\label{eq6}
}
Then instituting this equation into \Eq{soliton4_1}, there is
\eq{
    \frac{\dif^2\tphi}{\dif\teta^2}=\me^{\tphi}-\frac{\rM a^{\frac{1}{2}}}{\sqrt{\left(\rM a^{\frac{1}{2}}\right)^{2}-2\tphi}}
}
Next multiply $\dfrac{\dif\tphi}{\dif\teta}$ and integral over $\teta$
\eq{
    \frac{1}{2}\left(\frac{\dif\tphi}{\dif\teta}\right)^2=\me^{\tphi}+\rM a^{\frac{1}{2}}\sqrt{\left(\rM a^{\frac{1}{2}}\right)^{2}-2\tphi}+\left(\rM a^{\frac{1}{2}}+1\right).\label{eq7}
}
As discussed previous, $\rM a^{\frac{1}{2}}$ is number or variable that deviates slightly from the unit, so define $\varepsilon\equiv\rM a^{\frac{1}{2}}-1\ll1$.
Thus \Eq{eq7} rewrites
\eq{
    \left(\frac{\dif\tphi}{\dif\teta}\right)^2=\frac{2}{3}\tphi^{2}(3\varepsilon-\tphi).\label{eq8}
}
The solution to \Eq{eq8} is finally expressed as
\eq{
    \tphi&=3\varepsilon\; \mathrm{sech}^{2}\left[\left(\frac{\varepsilon}{2}\right)^{\frac{1}{2}}\left(\tx-\int^{\ttauu}\rM \dif\ttauu'\right)\right]\nonumber\\
         &\simeq3\varepsilon\; \mathrm{sech}^{2}\left[\left(\frac{\varepsilon}{2}\right)^{\frac{1}{2}}\left(\tx-\int^{\ttauu}a^{-1/2} \dif\ttauu'\right)\right],\label{tphi}
}
where $\mathrm{sech}(x)$ is the hyperbolic secant function.

\subsection{Korteweg-de Vires equation of density perturbation\label{KdV}}

Since $\varepsilon\equiv\rM a^{\frac{1}{2}}-1\ll1$, expand density ratio, electric potential and velocity as series of $\varepsilon$:
\begin{subequations}\label{series}
\begin{numcases}{}
    \tN=1+\varepsilon N^{(1)}+\varepsilon^2 N^{(2)}+\cdots,\\
    \tphi=\varepsilon \phi^{(1)}+\varepsilon^2\phi^{(2)}+\cdots,\\
    \tu=\varepsilon \tu^{(1)}+\varepsilon^2\tu^{(2)}+\cdots.
\end{numcases}
\end{subequations}
Define new quantities appeared in \Eq{tphi}
\eq{
    \xi\equiv\varepsilon^{1/2}\left(\tx-\int^{\ttauu}\rM \dif\ttauu'\right)=\varepsilon^{1/2}\left(\tx-\int^{\ttauu}a^{-1/2} \dif\ttauu'\right)
}
and
\eq{
    \eta=\varepsilon^{3/2}\int^{\ttauu}a^{-1/2} \dif\ttauu'.
}
Change the coordinate from $(\tx,\ttauu)$ to $(\xi,\eta)$
\eq{
    \frac{\partial}{\partial\tx}=\varepsilon^{1/2}\frac{\partial}{\partial\xi},\ \frac{\partial}{\partial\ttauu}=a^{-1/2}\varepsilon^{3/2}\frac{\partial}{\partial\eta}-
    a^{-1/2}\varepsilon^{1/2}\frac{\partial}{\partial\xi}.
}
Thus \Eq{soliton4} becomes
\eq{
\left\{\begin{array}{l}
    \varepsilon \dfrac{\partial^{2} \tphi^{\prime}}{\partial \xi^{2}}=\mathrm{e}^{\tphi}-\tN, \\
    a^{-1/2}\varepsilon \dfrac{\partial \tN}{\partial \eta}-a^{-1/2}\dfrac{\partial \tN}{\partial \xi}+\dfrac{1}{a}\dfrac{\partial(\tN \tu)}{\partial \xi}=0, \\
    a^{-1/2}\varepsilon \dfrac{\partial \tu}{\partial \eta}-a^{-1/2}\dfrac{\partial \tu}{\partial \xi}+\dfrac{\tu}{a}\dfrac{\partial \tu}{\partial xi}=-\dfrac{\partial \tphi}{\partial \xi}.
\end{array}\right.\label{KdV1}
}
Then the solution to the first order of \Eq{KdV1} reads
\eq{
    \phi^{(1)}=n^{(1)}=a^{-\frac{1}{2}}n^{(1)}. \label{KdV_1}
}
The equations to the order of $\varepsilon^2$ reads
\begin{subequations}\label{KdV_2}
\begin{numcases}{}
    \frac{\partial^2\phi^{(1)}}{\partial\xi^2}=\phi^{(2)}+\frac{1}{2}\left[\phi^{(1)}\right]^2-N^{(2)},\label{KdV_2_1}\\
    -a^{-1/2}\frac{n^{(2)}}{\partial\xi}+a^{-1/2}\frac{n^{(1)}}{\partial\eta}+\dfrac{1}{a}\dfrac{\partial(N^{(1)} u^{(1)})}{\partial \xi}+\dfrac{1}{a}\dfrac{u^{(2)}}{\partial \xi}=0,\label{KdV_2_2}\\
    -a^{-1/2}\frac{u^{(2)}}{\partial\xi}+a^{-1/2}\frac{u^{(1)}}{\partial\eta}+\dfrac{u^{(1)}}{a}\dfrac{\partial u^{(1)}}{\partial \xi}=-\frac{\partial\phi^{(2)}}{\partial\xi}.\label{KdV_2_3}
\end{numcases}
\end{subequations}
Eliminate $\phi^{(2)}$ by inserting \Eq{KdV_2_1} to \Eq{KdV_2_3} and replace $u^{(2)}$ and $n^{(2)}$ by applying \Eq{KdV_2_2}, then we have the first order of density perturbation $n^{(1)}$
\eq{
    \frac{\partial n^{(1)}}{\partial\eta}+n^{(1)}\frac{\partial n^{(1)}}{\partial\xi}+\frac{a^{1/2}}{2}\frac{\partial^3 n^{(1)}}{\partial\xi^3}=0. \label{KdV2}
}
For convenience, it's necessary to introduce variables $n^{(1)}=(a^{1/2}/2)^{1/3}\theta$ and $\xi=(a^{1/2}/2)^{1/3}\rho$, inserting which in \Eq{KdV2},
\eq{
    \frac{\partial\theta}{\partial\eta}+\theta\frac{\partial\theta}{\partial\rho}+\frac{\partial\theta}{\partial\rho^3}=0.\label{KdV_}
}
This is the famous Korteweg-de Vries (KdV) equation.

The profile of unimodal wave is the same as \Eq{tphi} plotted in Figs. \ref{soliton1} and \ref{soliton2}, implying a stationary density fluctuation propagates without dissipation.
Figs. \ref{soliton1} and \ref{soliton2} illustrate both the electric and energy density fluctuations since they comes from the pertubation of particles' number density.
This perturbation could be considered as the initial condition of the generation and evolution of galaxies since it would not disappear during and after photon recombination
due to the gravitational attraction in absence of optic dissipation. Besides the amplitude of such a fluctuation should not be sufficiently large because of propotion to $\varepsilon\ll1$
but enough to regard it as an appropriate starting collapsing condition \cite{Subramanian_2016}. It also illustrates the width of solitary wave's profile has a close value of galaxies' diameter.

\subsection{Numerical Simulation\label{simulation}}

\begin{figure}
  \centering
  \includegraphics[width=3.0in,height=2.7in]{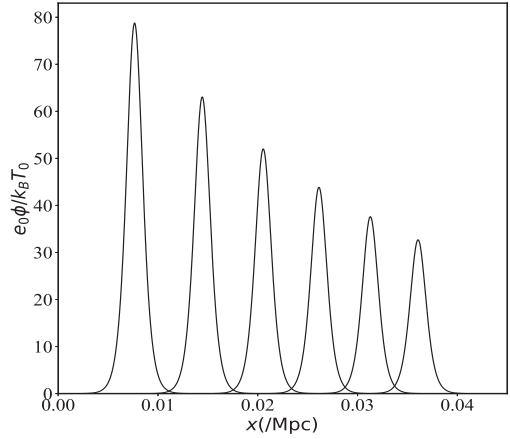}
  \caption{The profiles of solitary waves in \Eq{tphi} with $\varepsilon=0.01$ at the moments $\tau/\tau_*=$0.5, 0.6, 0.7, 0.8, 0.9 and 1.0 from left to right respectively.}
  \label{soliton1}
\end{figure}

\begin{figure}
  \centering
  \includegraphics[width=3.0in,height=2.7in]{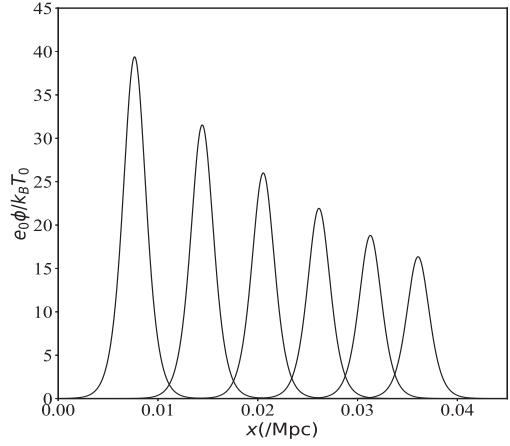}
  \caption{The profiles of solitary waves with $\varepsilon=0.005$.}
  \label{soliton2}
\end{figure}

As the calculations in Sec.~\ref{primordial}, the electric characteristic length evaluates $\tlambdae^{-1}=7.637\times10^{-5}\;\text{Mpc}^{-1}$.
If the major contributions are radiation and matters, the conformal time is briefly expressed
\eq{
    \tau=\frac{2}{\sqrt{\Omega_\text{m}H_0^2}}\Big[\sqrt{a+a_\text{eq}}-\sqrt{a_\text{eq}}\Big]. \label{tau}
}
Applying the recent Planck data, relevant cosmic data are best fitted as $a_*=1/1090$, $a_\text{eq}=1/3400$, where $a_\text{eq}$ and $a_*$
denote the scale factor at the present of radiation and matter equivalent and recombination. We have $c\tau_*=280.8\;\text{Mpc}$
and $c\tau_\text{eq}=112.9\;\text{Mpc}$ and $\tau_\text{eq}/\tau_*=0.4021$.
Then get the expression for scale factor as a function of conformal time via the inverse transform on \Eq{tau}
\eq{
    a&=\left[\dfrac{\tau\sqrt{\Omega_\text{m}H_0^2}}{2}+\sqrt{a_\text{eq}}\right]^2-a_\text{eq}\nonumber\\
     &=\left[\left(\sqrt{a+a_\text{eq}}-\sqrt{a_\text{eq}}\right)\eta+\sqrt{a_\text{eq}}\right]-a_\text{eq} \label{a}
}
with $\eta=\tau/\tau_*$. Thus, the formula in hyperbolic secant function becomes
\eq{
     &\tx-\int_{\ttauu_\text{eq}}^{\ttauu}\rM\dif\ttauu=\tx-\int_{\ttauu_\text{eq}}^{\ttauu}(1+\varepsilon)c_\textrm{s}\dif\ttauu\nonumber\\
    =&\tlambdae^{-1}\left[x-(1+\varepsilon)c\tau_*\sqrt{\frac{\kB T_0}{Mc^2}}\int_{\tau_\text{eq}}^\tau a^{-1/2}\dif(\tau/\tau_*) \right]\nonumber\\
    =&\tlambdae^{-1}\left[x-(1+\varepsilon)c\tau_*\sqrt{\frac{\kB T_0}{Mc^2}}\int_{\tau_\text{eq}/\tau_*}^\eta a^{-1/2}(\eta)\dif\eta \right].\label{xt}
}
Fig.~\ref{soliton1} and Fig.~\ref{soliton2} plot the solitary evolutions with $\varepsilon=0.01$ and $\varepsilon=0.005$. The differences between them appear only
the amplitude and width of the profile. It's notable such electric solitary waves vanish gradually while the electrons coupled with protons, but the density perturbation persists
still during the process because gravitation still works but optical pressure dissipates. The width of solitons' profile are around $D\sim0.01\;\text{Mpc}$,
which exactly approximate to the galaxies' diameters. So this may indicate the primordial nonlinear density perturbations after recombination promote the formation of galaxies.

Before the ending of this section, point needs to be explained. In Sec.~\ref{primordial} the primordial electric fields damp away by Landau damping, but in this section,
the fields propagate as solitons instead of attenuation. Are they paradoxical? Of course not. The former propagating as oscillating waves, or transverse wave precisely, these waves
absorb by plasma during propagating. While the latter being a solitary wave, or longitudinal wave precisely, these waves remain the stability during their propagations,
which is an intrinsic property inside the plasma matters.

\section{Acoustic Oscillations\label{ASS}}

Before the recombination epoch, the electrons and photons coupled tight with each other caused by Compton scattering, so the free path is much shorter than the Hubble horizon.

The unintegrated equations for proton and electron read
\eq{
    \frac{\dif f_\me(\textbf{x},\textbf{q},t)}{\dif t}=\langle C_{\me\rp}\rangle_{QQ'q'}+\langle C_{\me\gamma}\rangle_{pp'q'},\label{e}\\
    \frac{\dif f_\rp(\textbf{x},\textbf{Q},t)}{\dif t}=\langle C_{\me\rp}\rangle_{qq'Q'}+\langle C_{\rp\gamma}\rangle_{pp'Q'},\label{p}
}
where $C_{\me\rp}$ denotes the Coulomb scattering, $C_{\me\gamma}$ represents the Compton scattering between electron and photon which is inversely proportional to $m_\me^2$, and
$\langle \cdots\rangle_{QQ'q'}$ means the integral over momentum ${\bf Q}$, ${\bf Q},$ and ${\bf q}'$. Multiplying $\dfrac{\dif^3p}{(2\pi)^2}\dfrac{m_\me p\hp^i}{E}$ on both
sides of \Eq{e} and $\dfrac{\dif^3p}{(2\pi)^2}\dfrac{m_\rp p\hp^i}{E}$ on both sides of \Eq{p}, and integrating on momentum ${\bf q}$, it finally arrives at the baryonic velocity equation in Fourier space:
\eq{
     {v}_\rb'+Hv_b'+\mi k\Psi-\mi\frac{e_0}{m_\me}k\phi
    =\eta'\frac{4\rho_\mathrm{r}}{3\rho_\rb}\big[3\mi {\Theta}_1+v_\rb\big],\label{baryon_v}
}
where $v_\rb$ denotes the baryonic velocity, $\tilde{\Theta}_1$ denotes the dipole moment of perturbed photon's temperature, $\eta$ means the optic depth and
prime $'$ is the derivative to conformal time $\tau$. The Boltamann equations of photon for first two momentum give
\eq{
    \Theta_0'+k\Theta_1&=-\Phi',\label{theta0}\\
    \Theta_1'-\frac{k}{3}\Theta_0&=\frac{k\Psi}{3}+\mi\left[\Theta_1-\frac{\mi v_\rb}{3}\right].\label{theta1}
}

Insert \Eq{baryon_v} into itself and remain until the first order about baryonic velocity,
\eq{
    v_\rb=-3\mi\Theta_1+\frac{R}{\eta'}\left[v_\rb'+\frac{a'}{a}\nu_b+\mi k\Psi-\mi\frac{e_0}{m_e}k\phi\right]\label{baryon_v1}
}
with $R\equiv3\rho_\rb^{(0)}/4\rho_\mathrm{r}^{(0)}$. Then \Eq{theta1} becomes by inserting \Eq{baryon_v1}
\eq{
    \Theta_1-\frac{k}{3}=\frac{k\Psi}{3}+R\left(-\Theta_1'-\frac{a'}{a}\Theta_1+\frac{1}{3}k\Psi-\frac{1}{3}\frac{e_0}{m_\me}k\phi\right). \label{Theta1_1}
}
It could be rewritten as
\eq{
    \Theta_1'+\frac{a'}{a}\frac{R}{1+R}\Theta_1-\frac{k}{3(1+R)}\Theta_0=\frac{k}{3}\Psi-\frac{k}{3}\frac{e_0}{m_\me}\frac{\phi}{1+R}. \label{Theta1_2}
}
Eliminating $\Theta_1$ by inserting \Eq{theta0} into \Eq{Theta1_2}, we finally get the photons coupled with gravity containing the electric fields
\eq{
    \left[\frac{\dif^2}{\dif\tau^2}+\frac{R'}{1+R}\frac{\dif}{\dif\tau}+k^2c_\mathrm{s}^2\right]\left(\Theta_0+\Phi\right)
    = -\frac{k^{2}}{3}\Psi+\frac{k^{2}}{3(1+R)}(\Phi+\bphi),\label{baryon1}
}
with
\eq{
    c_\mathrm{s}=\sqrt{\frac{1}{3(1+R)}} \label{gcs}
}
and
\eq{
    \bphi=\frac{e_0\phi}{m_\me c^2}.
}
According to the Green's function theory, the photons' temperature coupled with gravity can be constructed if neglecting the damping term and considering only the oscillating term
\eq{
    &\Theta_0+\Phi=c_{1}s_{1}(\tau)+c_{2}s_{2}(\tau)        \nonumber     \\ &\quad+\frac{k^{2}}{3}\int_{0}^{\tau}[\Phi+\bphi-\Psi](\tau')\frac{s_1(\tau')s_2(\tau)-s_1(\tau)s_2(\tau')}{s_1(\tau')s'_2(\tau')-s'_1(\tau')s_2(\tau')}\dif\tau'\nonumber\\
    &=[\Theta_{0}(0)+\Phi(0)]\cos[kr_\text{s}(\tau)]     \nonumber     \\
    &\quad+\frac{k}{\sqrt{3}}\int_0^{\tau}[\Phi+\bphi-\Psi](\tau')\sin[k(r_\text{s}(\tau)-r_\text{s}(\tau'))]\dif\tau'.\label{as}
}
where $s_{1}(k,\tau)=\sin[kr_\text{s}(\tau)]\ \text{and}\ s_{2}(k,\tau)=\cos[kr_\text{s}(\tau)]$ are the two homogeneous solutions, and the sound horizon is defined as
\eq{
    r_\text{s}(\tau)=\int_{0}^{\tau}\dif\tau'c_\text{s}(\tau').\label{rs}
}
Compared with the condition in absence of electric field, \Eq{as} does not show any differences except a phase in terms of a sine integral
\eq{
    \Delta\varphi=\frac{k}{\sqrt{3}}\int_0^{\tau}\bphi(\tau')\sin[k(r_\text{s}(\tau)-r_\text{s}(\tau'))]\dif\tau'.
}

As discussed in Sec.~\ref{primordial} and Sec.~\ref{soliton}, the primordial electric fields damp out since Landau damping, so no need to calculate the influence on the acoustic oscillations.
However the intrinsic electric fields in plasma propagates as solitary waves, so it needs to evaluate the amplitude of this contribution.
$\tphi$ has been achieved in \Eq{tphi}, the temperature before recombination approximates $\kB T\sim0.1\text{eV}$, the energy of rest electron $m_\me c^2\sim0.5\text{MeV}$ and
the amplitude of soliton $\varepsilon\sim0.01$. Thus the additional term's amplitude approximates $\bphi\sim10^{-8}$. While the gravitational perturbation is
observed at order $\Phi,\ \Psi\sim10^{-5}$. So the electric fields merely increase the amplitude of acoustic oscillations by a order of $10^{-3}$, which means
the CMB's power spectra almost change nothing at acoustic peak. Besides, the electrical acoustic wave, appeared in \Eq{ecs}, propagates much slower than the
gravitational acoustic wave, appeared in \Eq{gcs}, the ratio between electrical acoustic horizon and gravitational acoustic horizon is
\eq{
    r=\frac{r_\mathrm{s}^{(\me)}}{r_\mathrm{s}^{(\text{g})}}=\frac{\int_{0}^{\tau}c_\mathrm{s}^{(\me)}\dif\tau'\;}{\int_{0}^{\tau}c_\mathrm{s}^{(\text{g})}\dif\tau'\;}
    \sim\sqrt{\frac{\kB T_0}{Mc^2}\cdot\frac{\Omega_\mathrm{b}h^2}{\Omega_\mathrm{r}h^2}}\sim10^{-4}.
}
This means the electrical acoustic peaks in CMB's power spectra locate at extremely large $\ell$'s, which roughly corresponds to the scale of galaxies or galaxy cluster,
consistent to the discussions in Sec.~\ref{soliton}.
\begin{figure}
  \centering
  \includegraphics[width=3.0in,height=2.7in]{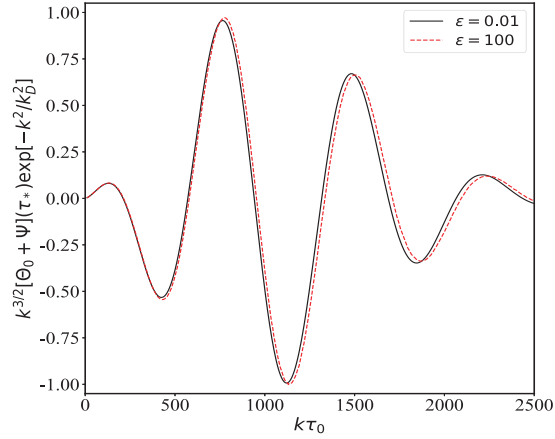}
  \caption{The monopole at recombination in a standard CDM model with $\varepsilon=0.01$ and $\varepsilon=100$.}
  \label{AS}
\end{figure}

The simulation on \Eq{baryon1} needs a further treatment on solitary solution in \eqref{tphi}.
Multiplying $\tau_*$ on both sides of \Eq{baryon1} and defining $\eta=\tau/\tau_*$, then it rewrites
\eq{
    &\left[\frac{\dif^2}{\dif\eta^2}+\frac{\dif R/\dif \eta}{1+R}\frac{\dif}{\dif\eta}+(\tau_*/\tau_0)^2(k\eta_0)^2c_\mathrm{s}^2\right]\left(\Theta_0+\Phi\right)
    = \nonumber\\
    &\quad\quad-\left(\frac{\tau_*}{\tau_0}\right)^2\frac{(k\tau_0)^{2}}{3}\Psi+\left(\frac{\tau_*}{\tau_0}\right)^2\frac{(k\tau_0)^{2}}{3(1+R)}(\Phi+\bphi),\label{baryon2}
}
here $\tau_0=14161.5\,\text{Mpc}$ denoting the conformal time at present time. Based on the Fourier transform
\eq{
    \mathcal{F}[\mathrm{sech}(x)]=k\pi\,\mathrm{scsh}\left(\dfrac{k\pi}{2}\right),
}
the transformation on $e_0\tphi/mc^2$ reads
\eq{
    \mathcal{F}[\bar{\phi}]=&\frac{\kB T_0}{mc^2a}\frac{3(2\varepsilon)^{1/2}\tlambdae}{c\tau_0}\nonumber\\
    &\times\exp\left(\mi\sqrt{\frac{\varepsilon}{2}}\frac{c\tau_*}{\tlambdae}\sqrt{\frac{\kB T_0}{Mc^2}}\int_0^\eta a^{-\frac{1}{2}}\dif\eta'\right)\nonumber\\
    &\times\pi kc\tau_0\,\mathrm{scsh}\left(\sqrt{\frac{2}{\varepsilon}}\frac{\pi \tlambdae }{2c\tau_0}kc\tau_0\right), \label{Fbphi}
}
where $\mathrm{scsh}(x)$ is the hyperbolic cosecant function. The Bardeen potentials follow the convenient fits \cite{Hu}:
\eq{
    \Phi(k,y)=\bar{\Phi}\left\{[1-T(k)]\exp[-0.11(ky/k_\text{eq})^{1.6}]+T(k)\right\},\\
    \Psi(k,y)=\bar{\Psi}\left\{[1-T(k)]\exp[-0.097(ky/k_\text{eq})^{1.6}]+T(k)\right\},
}
where $T(k)$ is the BBKS transfer function and $y\equiv a/a_\text{eq}$. Inserting the relevant equations into \Eq{baryon2}, then the numerical results are simulated in Fig.~\ref{AS}.
The black solid line represents the exact solution with $\varepsilon=0.01$ which almost totally coincides with the line of $\varepsilon=0$. While, the red dashed line represents the condition
with $\varepsilon=100$. This condition goes against the assumption $\varepsilon\ll1$, but it has already illustrated that the electric potential contributes a phase on the
CMB's power spectrum.

This section could be included in a brief statement: the CMB's power spectra could merely distinguish the contributions from primordial electric fields.

\section{\label{conclusion}Conclusions and further discussions}

In this paper, three problems are discussed: the damping of primordial electric fields, electric solitons and their effects on acoustic oscillations.
As calculated in relevant cosmological monographs \cite{2003moco.book.....D}, the effects on CMB from electromagnetic fields are often ignored,
although, in plasma, such effects always play extremely significant roles.

First, the evolutions of primordial electric fields are semi-precisely calculated, but the results are similarly with the ones in plasma physics,
which show the primordial fields dissipate out no matter at sub-horizon scale or at super-horizon scale by Landau damping effect, especially the fields
having entered the electric acoustic horizon. It means the effects of neutrino's recombination don't need take into account since the initial perturbations
have already damped out till the electron's recombination. In a word, there is no need to consider the electromagnetic initial conditions but just only consider
the gravitational initial conditions from inflation.
Second, I proof that the electric fields propagate as solitary waves instead of oscillating waves and their speed is much slower than baryonic acoustic oscillations.
And third, it illustrates the electric solitons merely affect the shape of CMB's power spectra. However, the protonic density fluctuation also shows a form of
solitary wave described by a KdV equation. Since proton has a much large mass than electron dose, such a nonlinear
perturbation grows continuously in the presence only of the gravitation, so it's regarded as the initial evolutionary condition of galaxies.

This paper illustrates the origin of electromagnetic fields after recombination, and offers an effective method to compute the electric influences on CMB's power spectra.
In \Eq{electrodynamics1}, the mechanic equation containing only the electric fields is obtained by setting the component $\mu=0$.
Similarly, the equation including the electric and magnetic fields could be obtained by setting $\mu=i$. It's not hard to imagine the
Vlasov equation of the second moment will introduce the higher order tensors, like magnetic fields, anisotropic tensor (or pressure tensor) and tensor perturbations.
Proper cutoff on higher moment will simplify the calculations.
But the computations will be quite complex and relevant results must be attractive.

\section*{Acknowledgment}

This project is supported by the National Natural Science Foundation of China (Grants No. 11864030) and
Scientific Research Funding Project for Introduced High Level Talents of IMNU (Grants No. 2020YJRC001).
I also thank my undergraduate students Jian-Kia Yin, Jia-Xin Ma, Jia-Xian Liu and Lu-Jie Shi help me input the formulas.

\appendix

\section{\label{Cauchy}Integral of Cauchy's type}

$F(x)$ defined in \Eq{Fx} could analytically calculated under two special conditions. The first is high frequency limit, i. e. $x\gg1$, then it approximates
\eq{
    F(x)&=\frac{x}{\sqrt{\pi}}\int^{+\infty}_{-\infty}\frac{\me^{-z^2}}{z-x-\mi0}\dif z           \nonumber\\
        &=\mathrm{P.\;V.}\frac{x}{\sqrt{\pi}}\int^{+\infty}_{-\infty}\frac{e^{-z^{2}}}{z-x}\dif z+\frac{x}{\sqrt{\pi}}\mi\pi \me^{-x^2} \nonumber\\
        &=\mathrm{P.\;V.}\frac{1}{\sqrt{\pi}}\int^{+\infty}_{-\infty}\frac{e^{-z^{2}}}{\frac{z}{x}-1}\dif z + \mi\sqrt{\pi}x \me^{-x^{2}} \nonumber\\
        &\simeq-\mathrm{P.\;V.}\frac{1}{\sqrt{\pi}}\int^{+\infty}_{-\infty}\dif z\;\me^{-z^2}\sum_{n=0}^\infty\left(\frac{z}{x}\right)^n+\mi\sqrt{\pi}x \me^{-x^{2}} \nonumber \\
        &\simeq -1-\frac{1}{2x^2}-\frac{3}{4x^4}+\mi\sqrt{\pi}x \me^{-x^2}, \label{Fxgg1}
}
where $\mathrm{P.\;V.}$ means the Cauchy's principal value \cite{principal}, and it has been employed two Gaussian integrals \cite{integral}:
\eq{
    \int_{-\infty}^{+\infty}x^{2n}\me^{-px^2}&\dif x=\frac{(2n-1)!!}{(2p)^n}\sqrt{\frac{\pi}{p}},\nonumber\\
             &(p>0,\quad n=0,\;1,\;\cdots) \label{Gauss1}
}
and
\eq{
    \int_{-\infty}^{+\infty}x^{2n+1}\me^{-px^2}\dif x=0,\ (p>1,\quad n=\cdots,\;-1,\;0,\;1,\;\cdots).\label{Gauss2}
}

The low frequency limit condition is evaluated as
\eq{
    F(x)&=\frac{x}{\sqrt{\pi}}\int_{-\infty}^{+\infty}\frac{\me^{-(z-x)^2}}{z-\mi0}\dif z \nonumber\\
        &=\mathrm{P.\;V.}\frac{x}{\sqrt{\pi}}\int_{-\infty}^{+\infty}\dif z\;\frac{\me^{-z^2}}{z}\left[1-2zx+\frac{1}{2!}(4z^2-2)x^2\right.\nonumber\\
        &\quad\quad\quad\quad\quad\quad\left.-\frac{1}{3!}(8z^3+12z)x^3+\cdots\right]+\mi\sqrt{\pi}x\nonumber\\
        &\simeq-2x^2+\frac{4}{3}x^4+\mi\sqrt{\pi}x,
}
where, of course, \Eq{Gauss1} and \eqref{Gauss2} have also been applied.

\end{document}